\documentclass[aps,prl, twocolumn, floatfix, showpacs]{revtex4-1}

\pdfoutput=1

\usepackage{graphicx}
\usepackage{latexsym}
\usepackage{amsmath}
\usepackage{graphics}
\usepackage{amssymb}
\usepackage{layout}
\usepackage{verbatim}
\usepackage{amsfonts,epsfig}
\usepackage{bm}
\usepackage{amsmath}

\newcommand{\up}{\uparrow}
\newcommand{\dn}{\downarrow}

\begin{document}

\title{Search for Majorana fermions in multiband semiconducting nanowires}
\author{Roman M.~Lutchyn$^{1,3}$}
\author{Tudor D. Stanescu$^{1,2}$}
\author{S. Das Sarma$^1$}
\affiliation{$^1$ Joint Quantum Institute and Condensed Matter Theory Center, Department of Physics,
University of  Maryland, College Park, Maryland 20742-4111, USA\\
$^2$ Department of Physics, West Virginia University, Morgantown, WV 26506, USA\\
$^3$Microsoft Research, Station Q, Elings Hall, University of California, Santa Barbara, CA 93106, USA}

\date{compiled \today}

\begin{abstract}

We study multiband semiconducting nanowires proximity-coupled with
an s-wave superconductor. We show that when odd number of subbands
are occupied the system realizes non-trivial topological state
supporting Majorana modes localized at the ends. We
study the topological quantum phase transition in this system and
analytically calculate the phase diagram as a function of the
chemical potential and magnetic field. Our key finding is
that multiband occupancy not only lifts the stringent constraint of
one-dimensionality but also allows to have higher carrier density in
the nanowire and as such multisubband nanowires are better-suited for observing the Majorana
particle. We study the robustness of the topological phase by including the effects of the short- and long-range disorder. We show that in the limit of strong interband mixing
there is an optimal regime in the phase diagram (``sweet spot")
where the topological state is to a large extent
insensitive to the presence of disorder.
\end{abstract}

\pacs{03.67.Lx, 71.10.Pm, 74.45.+c}

\maketitle

Looking for the elusive Majorana particles is one of
the most active and exciting current topics in all of
physics~\cite{Franz}.  Although originally proposed as a model
for neutrinos, the current search for Majorana particles is mostly
taking place in condensed matter or atomic
systems~\cite{dassarma_prl'05, Wimmer} where these mysterious particles,
which are their own anti-particles, emerge as effective quasiparticles from an
underlying fermionic Hamiltonian.  Quite apart from the intrinsic
interest associated with the exotic Majorana particles, the
possibility that they can be used in carrying out fault-tolerant
topological quantum computation~\cite{nayak_RevModPhys'08} by
suitably exploiting their non-Abelian braiding statistics gives an
additional technological impetus in the subject.  It has been known
for a while~\cite{kitaev'01, Fu_Kane_prb'09, lutchyn'10, oreg'10}
that, under suitable conditions, Majorana particles could exist at
the ends of 1D nanowires in the presence of the
appropriate superconducting (SC) pairing. Also, it has been
recently shown that the network of Majorana wires can be used for
braiding~\cite{alicea'10} and topological quantum
computation~\cite{sau_network}.  Although the semiconducting (SM)
nanowires~\cite{lutchyn'10, oreg'10} are promising candidates for
observing the Majorana, experimental realization of these proposals
is challenging because obtaining strictly 1D nanowires
is a very demanding materials problem~\cite{Doh'05}.  In this Letter we establish that one dimensionality, i.e. the occupancy of
one only subband in the nanowire,
is completely unnecessary, and Majorana particles can exist under
rather general and robust conditions even when several subbands are occupied in the nanowire.  More importantly, we prove the remarkable counter-intuitive result that
the multisubband system is, in fact, better-suited in observing the
Majorana than the strict 1D limit.
We carry out an analytic theory establishing our main
results and provide support for it by independent numerical
calculations. We also study the robustness of the topological phase against short- and long-range disorder and show that there is an optimal parameter regime where the system is most stable with respect to disorder. We believe that our results would go a long way in
providing the most suitable solid-state system for the eventual observation of
the Majorana particles.

\begin{figure}
\centering
\includegraphics[width=3.3in,angle=0]{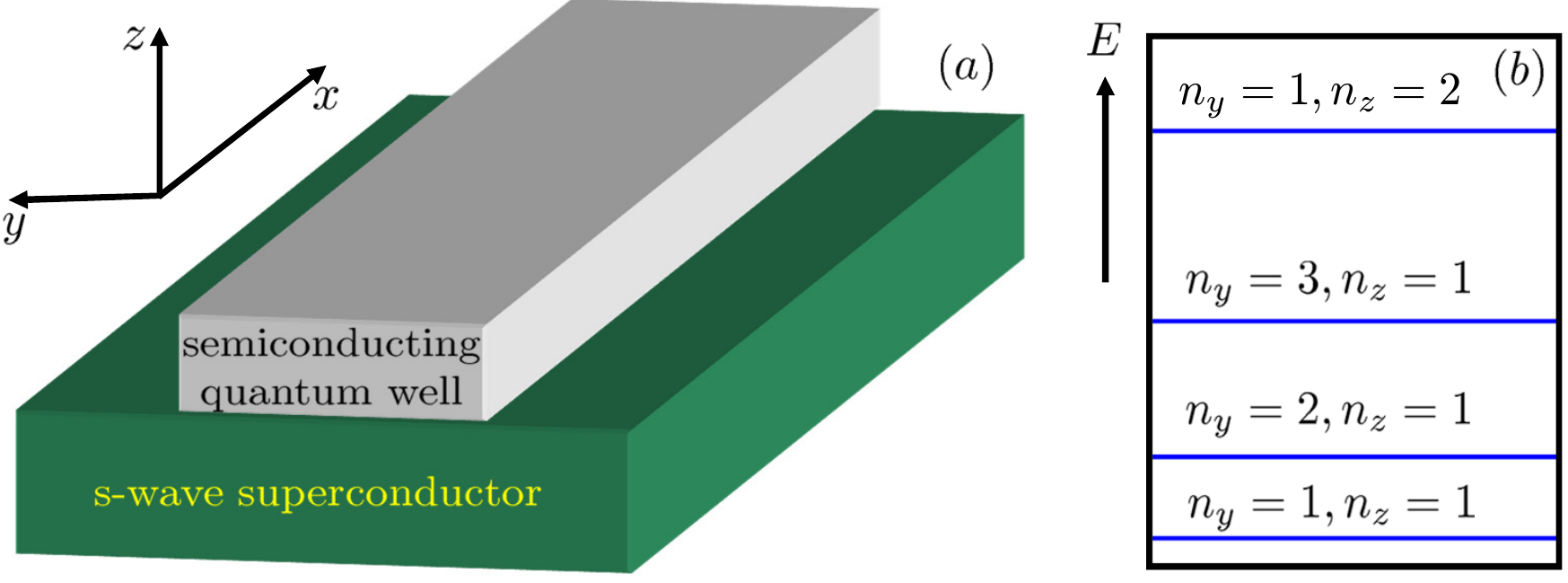}
\caption{(Color online) (a) Schematic plot of the quasi-1D nanowire proximity-coupled with an s-wave
SC. The rectangular quantum well has the dimensions
$L_z$, $L_y$ and $L_x$: $L_z \ll L_y \ll L_x$. The nanowire can be
top gated to control chemical potential in it. The method for
fabricating the proposed quantum well heterostructure
based on InAs has been demonstrated, see, e.g.,
Ref.~\cite{Takayanagi'85, mason'98}.(b) Schematic plot of the lowest energy
subbands due to the transverse confinement.}\label{fig:device}
\end{figure}

In this Letter we propose to study Majorana physics in
SM quantum well based on, for example, InAs-AlSb
heterostructure~\cite{mason'98}.  The active system consists of  a
SM with strong spin-orbit interaction proximity-coupled
with an s-wave SC, see
Fig.~\ref{fig:device}a. The rectangular quantum well has
the dimensions $L_z$, $L_y$ and $L_x$ as shown in
Fig~\ref{fig:device}a. We consider here the case of a strong confinement
in the $\hat{z}$ direction such that $L_z \ll L_y,L_x$ so that only the lowest subband with respect to
the $\hat z$-axis eigenstates is occupied. Then, the single-particle
Hamiltonian takes the usual form for the 2D SM in the
presence of the spin-orbit Rashba interaction ($\hbar=1$):
\begin{align}
{\cal H}_{\rm SM}&=\int dx dy \, \psi^\dag_{\sigma}(x,y)\hat H_{\sigma \sigma'}\psi_{\sigma'}(x,y)\label{eq:H0a}\\
H&=-\frac{\partial_x^2+\partial_y^2}{2m^*}-\mu-i\alpha (\sigma_x \partial_y-\sigma_y\partial_x)+V_x\sigma_x, \label{eq:H0b}
\end{align}
where $m^*$, $\alpha$ and $\mu$ are the effective mass, the strength
of spin-orbit interaction and chemical potential,
respectively. The latter can be controlled using the gate electrodes~\cite{Doh'05, mason'98}.
The last term in Eq.~\eqref{eq:H0b} corresponds to the Zeeman term
due to the applied external magnetic field aligned along the $\hat
x$-axis, $V_x=g_{\rm SM} \mu_B B_x/2$. Note that magnetic field is
essential here - it opens up a gap in the spectrum at $p_x=0$ and allows one to avoid fermion doubling which is
detrimental for the existence of Majorana fermions. Because of the
large g-factor in the SM $g_{\rm InAs} \lesssim 35$,
fairly small in-plane magnetic field $B_x \lesssim 1$T  opens up a
sizable gap in the spectrum without significantly disturbing the
SC.

We now include the size quantization along $\hat y$-direction assuming
that $L_y \ll L_x$. One can notice that Hamiltonian~\eqref{eq:H0b}
is separable in $x-y$ coordinates and the field operator can be
written as
\begin{align}\label{eq:psi}
\!\!\psi(x, y)\!=\!\sum_{p_x, n_y=1,2,...}\!\sqrt{\frac{2}{L_y L_x}}\!\sin\left(\frac{\pi n_y y}{L_y}\right)\!e^{ip_x x}\! a_{p_x, n_y},
\end{align}
where $a_{p_x, n_y}$ is electron annihilation operator in a state
$n_y$ having momentum $p_x$. The physical parameter regime we
consider here corresponds to the confinement energy along $y$-direction being larger than all the
relevant energy scales of the Hamiltonian~\eqref{eq:H0b} so that
there are only few lowest subbands occupied, see
Fig.~\ref{fig:device}b. This assumption actually corresponds to the typical experimental situation in InAs nanowires~\cite{private}. For $\mu \!<\! 2E_{\rm sb}$, where $E_{\rm sb}\!=\!3\pi^2/2 m^* L_y^2$ is the subband energy difference, one can project the
wavefunction to the lowest two subbands $n_y\!=\!1,2$ in \eqref{eq:psi} and simplify
the Hamiltonian~\eqref{eq:H0a}. By introducing the
spin-band spinors $\Phi\!=\!(c_{p_x\up}, c_{p_x\dn}, d_{p_x\up},
d_{p_x\dn})$ where the annihilation operators $c_{p_x}$ and
$d_{p_x}$ correspond to $a_{p_x, n_y\!=\!1}$ and $a_{p_x, n_y\!=\!2}$,
respectively, the single-body Hamiltonian becomes ${\cal H}=\sum_{p_x}
\Phi^\dag(p_x)H_{\rm red}\Phi(p_x)$ with $H_{\rm red}$ being defined
as
\begin{align}\label{eq:Hred}
\!\!\!H_{\rm red}\!\!=\!\!\frac{p_x^2}{2m}\!-\!\mu\!-\!\alpha\sigma_yp_x\!+\!E_{\rm sb}\!\frac{1\!-\!\rho_z}{2}\!-\!E_{\rm bm}\sigma_x\rho_y\!+\!V_x \sigma_x.
\end{align}
Here Pauli matrices $\sigma_i$ and $\rho_i$  act on the spin and
band degrees of freedom. The band mixing energy $E_{\rm bm}$
corresponds to the expectation value $\hat p_y$ operator between
different band eigenstates, i.e. $E_{\rm
bm}=\int_0^{L_y}dy\frac{2\alpha}{L}\sin(\frac{2\pi y}{L})\partial_y
\sin(\frac{\pi y}{L})=\frac{8\alpha}{3L_y}$.

We now study topological properties in this regime
with low number of subbands occupied. We investigate here whether Majorana fermions survive and are robust in this quasi-1D geometry. The multiband proximity-induced SC can be described as
\begin{align}\label{eq:HSC}
H_{\rm SC}=\sum_{p_x}&\left[\Delta_{11}c^\dag_{p_x\up}c^\dag_{-p_x\dn}+\Delta_{22}d^\dag_{p_x\up}d^\dag_{-p_x\dn}\right.  \\
&\left.+\Delta_{12}d^\dag_{p_x\up}c^\dag_{-p_x\dn}+ \Delta_{12}  c^\dag_{p_x\up}d^\dag_{-p_x\dn}+h.c.\right]\nonumber,
\end{align}
where the induced SC pairing potentials $\Delta_{ij}$
depend on the microscopic details of the interface between
SM and SC, \emph{e.g.} rough or smooth
interface. In the former case the magnitude of $\Delta_{12}$ can be a sizable fraction of $\Delta_{11}$. Taking into account the total Hamiltonian $H_{\rm tot}\!=\!H_{\rm red}\!+\!H_{\rm SC}$ we can now define the Nambu spinor as follows: $\Psi(p)\!=\!\left(c_{p_x\up},
c_{p_x\dn}, d_{p_x\up}, d_{p_x\dn},
c^\dag_{-p_x\up},c_{-p_x\dn}^\dag, d_{-p_x\up}^\dag,
d^\dag_{-p_x\dn}\right)^T$. In this convention for Nambu spinors the
BdG Hamiltonian for two subband model reads
\begin{align}\label{eq:BdG}
&\!\!H_{\rm BdG}(p_x)\!=\!\!\left(\!\frac{p_x^2}{2m^*}\!-\!\mu\!-\!\sigma_yp_x\!+\!\frac{E_{\rm sb}}{2}(1\!-\!\rho_z)\!+\!V_x\sigma_x\!\right)\!\tau_z\!  \\
\!\!&\!\!-\!\!E_{\rm bm}\sigma_x\rho_y\!+\!i\sigma_y\! \left[\rho_x |\Delta_{12}|\!+\!\Delta_{+}\!+\!\rho_z\Delta_{-}\!\right]\!\!(i\tau_y\cos \varphi\!+\!i\tau_x\sin \varphi)\nonumber,
\end{align}
where Pauli matrices $\sigma_i$, $\rho_i$ and $\tau_i$ act on
spin, band and Nambu degrees of freedom of the spinor $\Psi(p)$,
respectively; $\Delta_{\pm}\!=\!(|\Delta_{11}|\!\pm\!|\Delta_{22}|)/2$ and $\varphi$ is the SC phase. The particle-hole symmetry for $H_{\rm BdG}$~\eqref{eq:BdG} reads
$\Theta H_{\rm BdG}(p)\Theta^{-1}=-H_{\rm BdG}(-p)$,
where $\Theta$ is an anti-unitary operator $\Theta\!=\!\tau_x K$ with
$K$ denoting the complex conjugation.

The presence of Majorana modes in the system and the corresponding phase diagram can be obtained using topological arguments due to Kitaev~\cite{kitaev'01}. Following Ref.~\cite{kitaev'01} we introduce $Z_2$ topological index $\cal M$ (Majorana number):
\begin{align}\label{eq:Majorana}
{\cal M}={\rm sgn}[{\rm Pf} B(0)]{\rm sgn}[{\rm Pf} B(\pi/a)]=\pm 1,
\end{align}
where $\pm 1$ corresponds to topologically trivial/non-trivial states. Here antisymmetric matrix $B$
defines the Hamiltonian of the system in the Majorana basis~\cite{kitaev'01}. Rather
than computing the transformation matrix to the Majorana basis as was originally done in Ref.~\cite{kitaev'01}, we note following
Refs.~\cite{lutchyn'10, Ghosh'10} that the antisymmetric matrix
$B(p_x)$ can be simply constructed by the virtue of the
particle-hole symmetry. Indeed, the matrix $B(P)\!=\!H_{\rm BdG}(P)\tau_x$ needs to be calculated at the particle-hole invariant points where $H_{\rm
BdG}(P)\!=\!H_{\rm BdG}(-P)$ and $B$ is antisymmetric $B^T(P)\!=\!-B(P)$. In 1D there are two such points: $P\!=\!0,\!\frac{\pi}{a}$ with $\frac{\pi}{a}$ being the momentum at the end of the Brillouin zone and $a$ being the lattice spacing. (For the continuum
model considered here $\pi/a\! \rightarrow\! \infty$.) The function $\rm Pf$ in Eq.~\eqref{eq:Majorana} denotes Pfaffian of the antisymmetric matrix $B$. The computation of Pfaffian at
$P\!=\!\pi/a\!\rightarrow\! \infty$ is straightforward yielding ${\rm
sgn}[{\rm Pf} B(\pi/a)]\!=\!+1$.
\begin{figure}
\centering
\includegraphics[width=3.3in,angle=0]{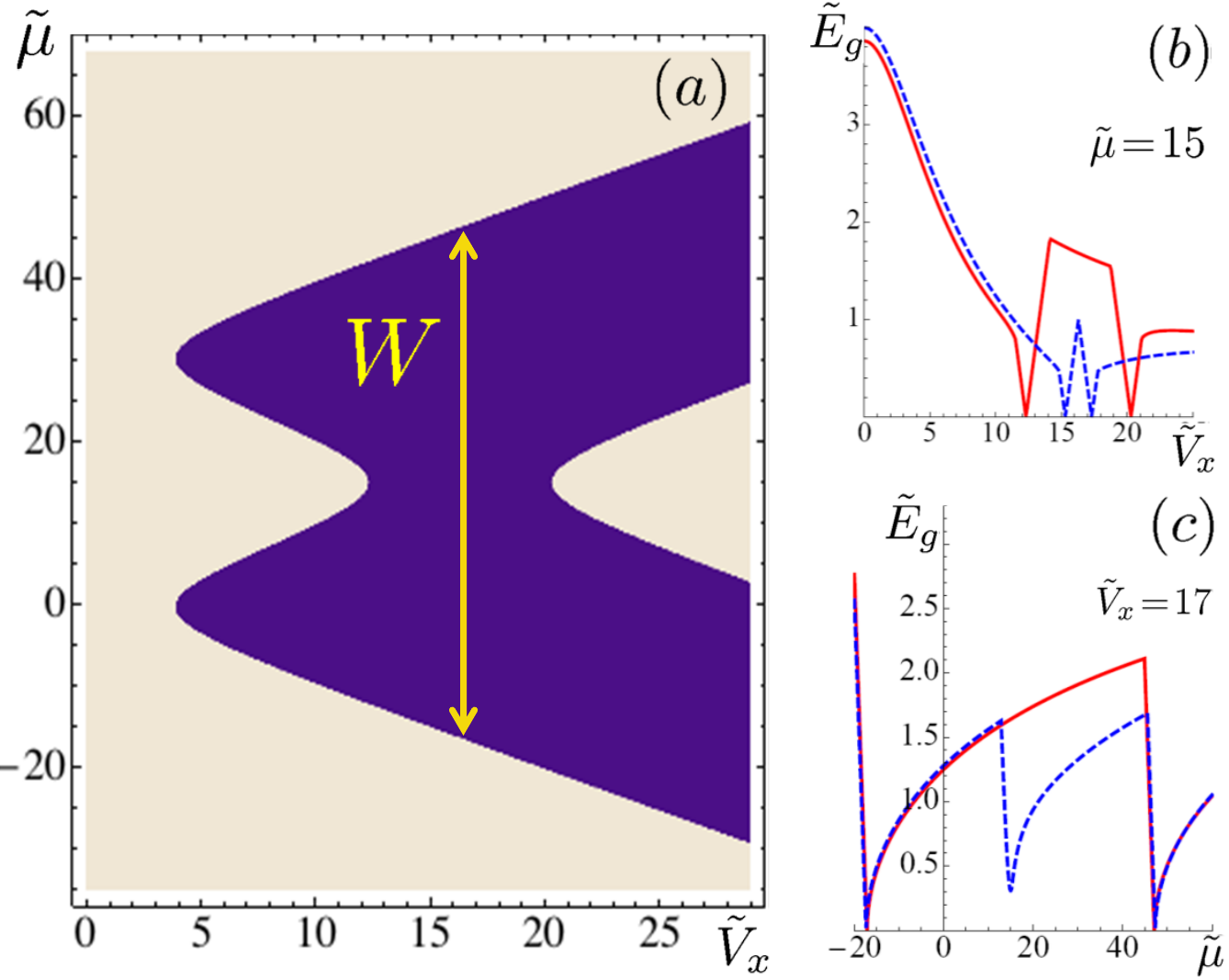}
\caption{(Color online) (a) Phase diagram for the two band
 nanowire model as a function of the chemical
potential $\tilde \mu$ and external magnetic field $\tilde V_x$. The width of the topological region is largest at the ``sweet" spot: $\tilde W\!\approx\!60$. Here tilde denotes re-scaled energy $\tilde E\!=\!E/m^*\alpha^2$ and $\tilde \Delta_{12}\!=\!4$. The
light and dark regions correspond to topologically trivial/non-trivial phases. (b) and (c) Quasiparticle excitation gap obtained using Eq.~\eqref{eq:BdG} as a function of $\tilde V_x$ and $\tilde \mu$. The solid (red) and dashed (blue) lines correspond to $\tilde \Delta_{12}\!=\!4$ and $\tilde \Delta_{12}\!=\!1$, respectively. The closing of the gap for $\tilde \Delta_{12}\!=\!4$ (solid red line) is consistent with the phase diagram shown in (a). The quasiparticle excitation gap at the ``sweet spot" strongly depends on the magnitude of $\Delta_{12}$. We assume here
$m^*=0.04m_e$ with $m_e$ being electron mass and $\alpha=0.1$eV$\AA$
yielding $m^*\alpha^2\approx 0.6$K. We used realistic
parameters $L_y=130$nm and $\tilde E_{\rm sb}=30$, $\tilde E_{\rm
bm}=5$ and $\tilde \Delta_{11}=\tilde \Delta_{22}=4$. }\label{fig:phase_diag}
\end{figure}
Thus, the phase boundary for the transition between topologically trivial and
non-trivial phases is given by the sign change of  ${\rm Pf} B(0)$
which can only happen when the bulk quasiparticle gap becomes zero,
i.e ${\rm Det} \, H_{\rm BdG}(P)\!=\!0$, see Fig.~\ref{fig:phase_diag}. This is a generic phenomenon since
the topological reconstruction of the fermionic  spectrum cannot
occur adiabatically and requires the nullification of the bulk
excitation gap~\cite{read_prb'00, Nishida}. For a two-band model
${\rm Pf} B(0)$  can be calculated analytically
\begin{align}
{\rm Pf} B(0)\!&=\!\left(V_x^2\!-\!E_{\rm bm}^2\!+\!|\Delta_{12}|^2\!+\!\Delta_{-}^2\!-\!\Delta_{+}^2\!-\!E_{\rm sb} \mu \!+\!\mu ^2\right)^2\nonumber\\
&\!-\!V_x^2 \left(E_{\rm sb}^2\!-\!4 E_{\rm sb} \mu \!+\!4 \left(\Delta_{12}^2\!+\!\Delta_{-}^2\!+\!\mu ^2\right)\right)\nonumber\\
&\!+\!(E_{\rm sb} (\Delta_{-}\!+\!\Delta_{+})\!-\!2\Delta_{+}\mu )^2
\end{align}
allowing one to compute $\cal M$ as a
function of the physical parameters. The phase diagram showing a
sequence of topological phase transitions for the two subband nanowire is shown in Fig.~\ref{fig:phase_diag}a. We
now analyze the phase diagram in various regimes. In the limit $\mu,
|V_x|\! \ll \! E_{\rm sb}$ we find that ${\rm Pf} B(0)\!\approx\!-\!V_x^2 \!+\!
\Delta_{11}^2 \!+\!\mu^2$ recovering the previous results obtained for
the single band~\cite{lutchyn'10, oreg'10}. When $|V_x|\! \ll \! \mu\! \sim\!
E_{\rm sb}$ we find that ${\rm Pf} B(0)\!\approx\!-V_x^2\! +\! \Delta_{22}^2
\!+\!(E_{\rm sb}\!-\!\mu)^2$. Thus, the system supports Majorana modes as
long as $|V_x|\!>\!\sqrt{\Delta_{22}^2 \!+\!(E_{\rm sb}\!-\!\mu)^2}$. These results
can be intuitively understood within weak-coupling approximation since in both cases the Fermi level
crosses odd number of bands in the interval $(0, \frac{\pi}{a})$. The most
interesting parameter regime is $\mu\! \sim\! E_{\rm sb}/2$ which
corresponds to the ``sweet spot" in the phase diagram, see Fig.~\ref{fig:phase_diag}a. At this point
the system is to a large extent insensitive to chemical potential
fluctuations and, thus, this regime provides a promising route to realizing a
robust topological SC phase. At $\mu=E_{\rm sb}/2$ the width of the
topologically non-trivial region is given by $E_{\rm
sb}/2-\Delta_{12}< |V_x| < E_{\rm sb}/2+\Delta_{12}$ to a leading
order in $1/E_{\rm sb}$. This is non-perturbative result and the
SC state emerging here is determined by the strong
interband mixing due to $\Delta_{12}$. The presence of a sizable $\Delta_{12}$ is crucial for the topological stability of  the non-trivial SC phase and the magnitude of the quasiparticle excitation gap at the ``sweet spot" strongly depends on the value of $\Delta_{12}$, see Figs.~\ref{fig:phase_diag}b and c.

\begin{figure}
\centering
\includegraphics[width=3.42in,angle=0]{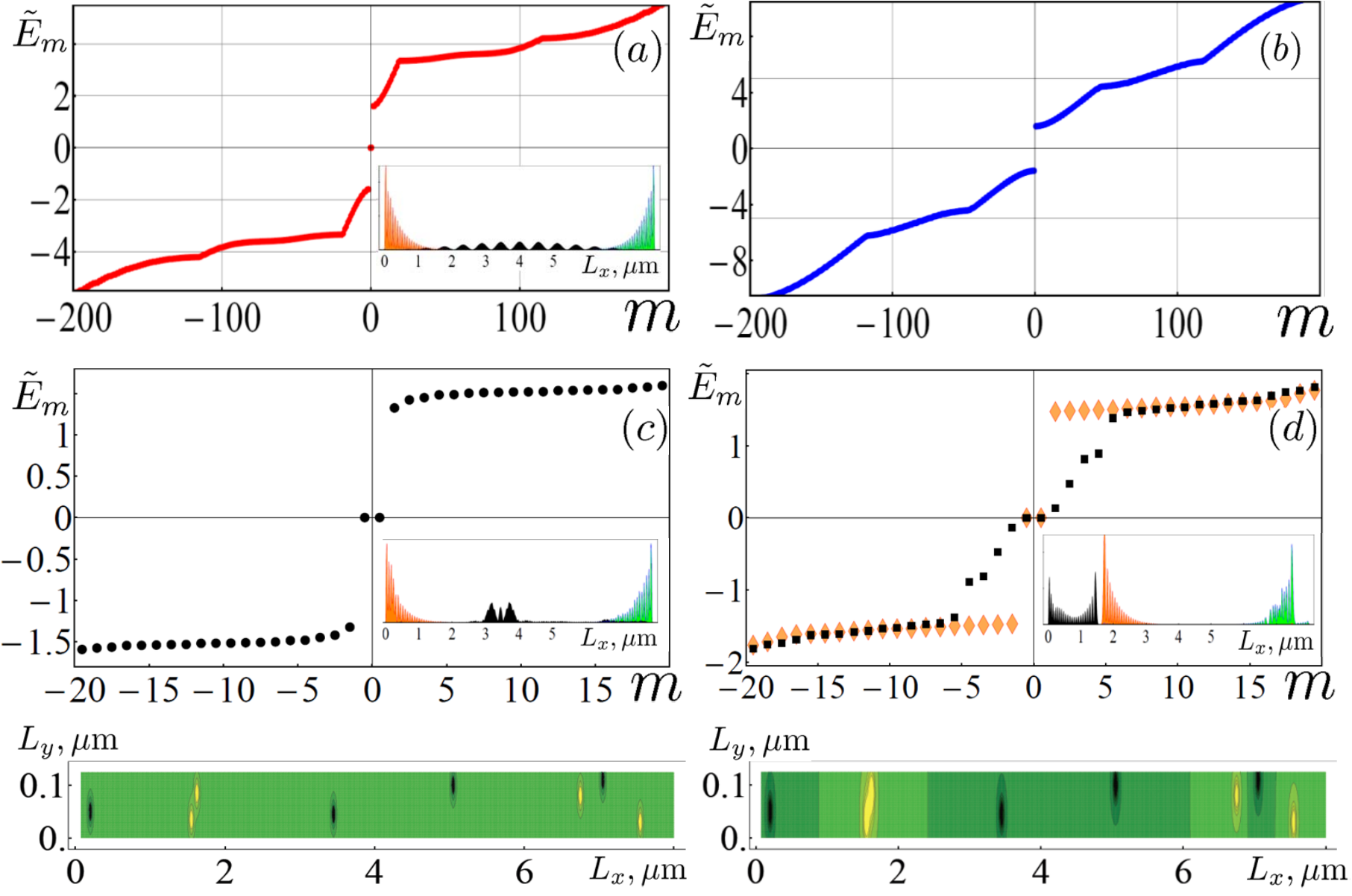}
\caption{(Color online) (a) and (b) Energy spectrum $\tilde E_m$ for a finite-size nanowire obtained by numerical diagonalization of $H_{\rm tot}\!=\!H_{\rm SM}\!+\!H_{\rm SC}$
for $\tilde \mu=15$, $\tilde V_x\!=\!15$ and $\tilde V_x \!\approx \!8$,
respectively. Here $\tilde E\!=\!E/m^*\alpha^2$ and $m$ labels eigenvalues of $H_{\rm tot}$. Inset: lowest lying energy states. Majorana zero-energy modes are present in (a)
and disappear in (b).
(c) and (d) Energy spectra for a given disorder realization with
impurity potentials shown at the bottom of the panels (c) and (d) corresponding to $\lambda\!=\!16$nm and $\lambda\!=\!260$nm, respectively. The dark (light) colors denote positive (negative) $U_{0j}$. c) The spectrum at the ``sweet spot" for the short-range disorder shown at the bottom ($\tilde U_0\!=\!100$). Inset: lowest-lying eigenstates. The topological phase is robust against short-range disorder, {\it i.e.} the disorder affects extended states but leaves Majorana modes intact.
d) The spectrum at the ``sweet spot" with long-range disorder for $\tilde U_0\!=\!100$ (squares) and $\tilde U_0\!=\!25$ (diamonds). The impurity potential $U_{\rm imp}$ is shown at the bottom of the panel. Inset: lowest-lying eigenstates for $\tilde U_0\!=\!100$. The topological phase is stable as long as $U_0\! < \! W/2$. For $\tilde U_0\!=\!100$ additional Majorana modes are localized at the impurities and the topological phase collapses by splitting into fragments of topological and non-topological regions, see inset. Both types of disorder lead to the emergence of the additional subgap states localized at the ends. Here we used parameters specified in Fig.~\ref{fig:phase_diag}.
}\label{fig:qp_spectrum}
\end{figure}

In order to establish the robustness of the topologically
non-trivial phase near the ``sweet spot" we have done independent
numerical simulations for a finite multiband nanowire with $L_x\! \!\sim\!\! 10\mu$m and $L_y\!\sim\!0.1\mu$m. The results obtained by
numerical diagonalization of the real-space Hamiltonian $H_{\rm
tot}\!=\!H_{\rm SM}\!+\!H_{\rm SC}$ are shown in
Fig.~\ref{fig:qp_spectrum}a and b. One can notice that at the
``sweet spot" ($\tilde V_x\!\approx\!15$ and $\tilde \mu\!\approx\!15$) there is a pair of Majorana zero-energy states whereas
for a smaller magnetic field ($\tilde V_x\!\approx\! 8$) corresponding to the trivial phase the zero energy modes disappear corroborating the phase diagram
shown in Fig.~\ref{fig:phase_diag}a. Furthermore, at the ``sweet
spot" the zero energy states are well-separated from the continuum. Indeed, as shown in Fig.~\ref{fig:qp_spectrum}a the minigap $E_{\rm mn}$ constitutes a
sizeable fraction of the induced SC gap, $E_{\rm mn}\!\!\sim\!\!1$K. Thus, Majorana modes in quasi-1D nanowires are very robust against
thermal fluctuations which makes these systems very advantageous for the topological quantum computation. We also studied the robustness of the topological phase against disorder by adding the impurity potential $U_{\rm imp}(\bm r) \!=\! \sum_{j}\!  U_{0j} \frac{\exp(-|\bm r \!-\! \bm r_j|/\lambda)}{1\!+\!|\bm r\!-\!\bm r_j|/d}$ to the Hamiltonian $H_{\rm tot}$. Here $\bm r$ is a vector in $x\!-\!y$ plane, $r_j$ are random positions of the impurities, $\lambda$ is the screening length, $U_{0j}\!=\!\pm U_0$ is the impurity potential with random sign but constant magnitude $U_0$, and  $d$ is the cutoff regularizing 1/r potential at short distances. We considered here two types of disorder mimicking short-range impurities ($\lambda\!=\!16$nm) and long-range chemical potential fluctuations ($\lambda\!=\!260$nm), see Fig.~\ref{fig:qp_spectrum}c,d. In the former case, the topological phase is very robust against disorder even if $|U_0| \gg E_{\rm sb}$, see Fig.~\ref{fig:qp_spectrum}c. This can be qualitatively understood as follows: at a given position the disorder potential leads to a formation of two Majorana modes localized at the impurity. Because these Majorana states are close to each other they hybridize and form conventional subgap states and do not affect Majorana modes at the ends even if the impurity is fairly close to the edge, see Fig.~3c. On the other hand, the long-range disorder is more dangerous. In Fig.~\ref{fig:qp_spectrum}d we show energy spectrum for two cases: $U_0$ smaller ($\tilde U_0\!=\!25$) and larger ($\tilde U_0\!=\!100$) than $W/2$, see Fig.~\ref{fig:phase_diag}a.
For $U_0<W/2$, the topological phase is stable, i.e. the disorder can suppress excitation gap but does not affect Majorana modes. On the other hand, if $U_0 \!>\! W/2$, the disorder effectively creates inhomogeneous wire with many topological and non-topological regions, see Fig.~\ref{fig:qp_spectrum}d, i.e. each topological segment now becomes much smaller allowing for strong mixing of the Majorana modes at the opposite ends. Thus, our simulations explicitly demonstrate the importance of working at the ``sweet spot" where the width of the topological region $W$ is maximized and the topological phase is most robust against long-range disorder.

To conclude, we have derived the topological phase diagram for the
existence of Majorana particles in a realistic
quasi-1D semiconductor wire in the presence of multisubband occupancy.
Unexpectedly, we find robust and experimentally feasible "sweet spots"
in the chemical potential- Zeeman splitting phase diagram where
Majorana modes should stabilize at the ends of the wire.  The great
advantages of our proposed structure in detecting Majorana particles
are (i) its materials flexibility (i.e. no need to impose one
dimensionality or single channel constraint), and (ii) its immunity to
density (or chemical potential) fluctuations and disorder. The calculation of the energy spectrum for realistic experimental settings
suggests the possibility to test our theoretical results using local tunneling experiments, see Ref.~\cite{sau_long}. Tunneling of electrons to the
ends of the nanowire would reveal a pronounced zero-bias peak when
the system is in topologically non-trivial phase. This zero bias
peak will disappear in the trivial phase.

{\it Note added.} While finishing this manuscript we became aware of
Refs.~\cite{Wimmer, Potter} where multichannel generalization of the
spinless p-wave SC state was studied.

This work is supported by DARPA-QuEST and JQI-NSF-PFC.

\end{document}